# Imaging the granular structure of high-$T_c$ superconductivity in underdoped $Bi_2Sr_2CaCu_2O_{8+d}$

K.M. Lang*, V. Madhavan*, J.E. Hoffman*, E.W. Hudson*†‡, H. Eisaki§¦, S. Uchida§ & J.C. Davis*‡

* Department of Physics, University of California, Berkeley, CA 94720, USA

† Department of Physics, Massachusetts Institute of Technology, Cambridge, MA 02139-4301, USA

‡ Materials Science Division, Lawrence Berkeley National Laboratory, Berkeley, CA 94720, USA

§ Department of Superconductivity, University of Tokyo, Yayoi, 2-11-16 Bunkyoku, Tokyo 113-8656, Japan

¦ Department of Applied Physics, Stanford University, Stanford, CA 94205-4060, USA

**Granular superconductivity occurs when microscopic superconducting grains are separated by non-superconducting regions through which they communicate by Josephson tunneling to establish the macroscopic superconducting state[1]. Although crystals of the cuprate high-$T_c$ superconductors are not granular in a structural sense, theory indicates that at low hole densities the holes can become concentrated at some locations resulting in hole-rich superconducting domains[2-5]. Granular superconductivity due to Josephson tunneling through 'undoped' regions between such domains would represent a new paradigm for the underdoped cuprates. Here we report studies of the spatial interrelationships between STM tunneling spectra in underdoped $Bi_2Sr_2CaCu_2O_{8+d}$. They reveal an apparent spatial segregation of the electronic structure into ~3nm diameter domains (with superconducting characteristics and local energy gap D<50 meV) in an electronically distinct background. To explore whether this represents nanoscale segregation of two distinct electronic phases, we employ scattering-resonances at Ni impurity atoms[6] as 'markers' for the local existence of superconductivity[7-9]. No Ni-resonances are detected in any regions where D>50±2.5 meV. These observations suggest that underdoped $Bi_2Sr_2CaCu_2O_{8+d}$ is a mixture of two different short-range electronic orders with the long-range characteristics of a granular superconductor.**

Evidence from scanning tunneling microscopy (STM) for nanoscale spatial variations in the electronic characteristics of $Bi_2Sr_2CaCu_2O_{8+\delta}$ (Bi-2212) has been steadily accumulating[10-12]. Recent advances include observations of nanoscale regions with 'superconducting' spectra coexisting with regions characterized by 'pseudogap-like' spectra[13,14], and the observation of spatial inhomogeneities in the gap magnitude which correlate with those in integrated differential tunneling conductance[15]. Several new theoretical models have emerged in response to these data[16-22]. A frequent theme among these models is that strong electrostatic-potential variations are experienced by the itinerant holes. Such potential fluctuations could be caused by frustrated electronic phase separation[2,4,5], by unscreened dopant-atoms[15,19-21], or by some other unrelated disorder[16-18,22]. In any of these scenarios, the holes would redistribute themselves to minimize their total energy in the disordered potential landscape. At low hole-densities some regions would therefore remain doped while others are relatively undoped. Thus, two types of electronic order, superconducting and non-superconducting respectively, would coexist in different nanoscale regions of the same crystal [16-19]. If Josephson tunneling couples the superconducting domains, the crystal would exhibit granular superconductivity[13,23].

STM can be used to explore these predictions since it can access real-space electronic structure with atomic resolution. Here we report high-resolution STM studies of underdoped Bi-2212. The samples are single crystals grown by the floating zone method. The 'as-grown' samples have a hole-dopant level p≈0.18±0.02 while the underdoped samples are oxygen-depleted yielding $T_c$≈79K and p≈0.14±0.02. They are cleaved in cryogenic ultra-high vacuum at T<30K and, when inserted in to the STM head, show atomic resolution on the BiO surface plane. Our primary datasets consist of differential tunneling conductance spectra (dI/dV vs. V) measured at all locations in a given field-of-view. These *spectral-surveys* are essentially detailed spatial maps of the local density of electronic states (*LDOS*) as a function of energy E because $LDOS(E = eV) \propto dI/dV(V)$ where V is the sample bias.

Spectral-surveys can be analyzed to yield maps of the energy gap as a function of location (*gapmaps*) by defining Δ for each *dI/dV* spectrum as:

$$\Delta \equiv \frac{\Delta_+ - \Delta_-}{2} \qquad (1)$$

Here $\Delta_{+(-)}$ is the energy of the first peak above(below) the Fermi level. Peaks from scattering-resonances are disregarded, and Δ does not necessarily represent a *superconducting* energy gap.

In Figure 1 we show gapmaps extracted from two such spectral-surveys, each measured on a 560Å square area. They use an identical color scale. Fig. 1a is typical of gapmaps on underdoped Bi-2212, while Fig. 1b is typical of the slightly overdoped 'as-grown' samples. They are obviously quite different in appearance. The as-grown gapmap is dominated by large interconnected areas of low-Δ (red and yellow), interspersed with filamentary high-Δ areas (blue and black). This is in contrast to the underdoped sample in Fig. 1a in which the compact, spatially distinct, low-Δ regions (red and yellow) are surrounded by interconnected high-Δ regions (blue and black). For the as-grown sample shown, the mean gap value is $\overline{\Delta}$=35.6meV with σ=7.7meV, while for the underdoped sample shown $\overline{\Delta}$=50meV and σ=8.6meV. Were these gapmaps the only information available, one might suspect that reducing the oxygen-doping has merely shifted $\overline{\Delta}$. However, as we show below using both additional information from the spectral-surveys and new experimental techniques, the differences between as-grown and underdoped samples are more profound.

Even though the spectral-surveys yielding Fig. 1 had spatial resolution of ≈4Å, the spectra can show strong variations from one pixel to the next. Therefore, an accurate description of the electronic structure requires higher spatial resolution spectral-surveys. Fig. 2a shows a typical gapmap from such a high-resolution spectral-survey acquired with 128x128 pixels on the 147Å square area indicated by the white box in Fig. 1a. Figure 2b is a map of the gap-edge peak-amplitude, G(Δ−), from the same spectral-survey. In a conventional superconductor this would be a map of the coherence-peak heights. We use the same color bar to represent Δ in Fig. 2a and G(Δ) in 2b to illustrate the spatial correlations between these two different observables. In these figures one sees what appear to be compact, almost circular, domains. They are characterized by a low value of Δ, and a G(Δ) that rises rapidly from the domain edge to reach a maximum at the domain-center. We will refer to these regions as ***a**-domains*. As an example, a single α-domain is identified inside the white circle on Figs. 2a&2b. Different α-domains have different characteristic values of Δ, and multiple α-domains are clumped together at some locations. Intervening between the α-domains are percolative regions characterized by high Δ and low almost-constant G(Δ). We will refer to these spectroscopically distinct areas as ***b**-regions*.

Figures 2c&2d show the evolution of Δ and G(Δ) respectively with distance along the white line in Figs. 2a&2b. Here we see in detail how, within each α-domain, the value of Δ varies little, while G(Δ) rises rapidly from the perimeter reaching a maximum in the center. In the surrounding interconnected β-regions, Δ varies slowly while G(Δ) is low and almost constant. Figure 2 focuses on the typical evolution of Δ and G(Δ), but these parameters represent only a small subset of the information in a spectral-survey. To see directly how the electronic phenomena reported here are manifest in the raw data, we show in Fig. 3 the unprocessed dI/dV spectra measured along the white line in Figs. 2a&2b. The spectral evolution with passage through three α-domains (red) and the intervening β-regions (blue) can be seen.

Figures 2&3 demonstrate how on the basis of two observables, Δ and G(Δ), the α-domains and β-regions appear spectroscopically distinct. We refer to this situation as electronic *segregation*. It becomes apparent only in underdoped samples because Δ<50meV for the α-domains and these samples have Δ>50meV for ~50% of their area, while the as-grown samples studied have Δ>50meV for only ~10% of theirs. This apparent electronic segregation motivates a new picture of the effects of reducing oxygen-doping in Bi-2212.

The α-domains have characteristics usually associated with superconductivity, e.g. sharp gap-edge peaks indicating well-defined Bogoliubov quasiparticles and an increasing areal density associated with rising $T_c$. In contrast, the β-region spectra exhibit characteristics often associated with the pseudogap phase (although we cannot definitively identify these regions as such). Consequently, the emerging picture is reminiscent of a granular superconductor with coupled superconducting grains embedded in a distinct electronic environment.

As STM is a surface probe, we next consider whether the observed phenomena represent a property of bulk underdoped Bi-2212. Evidence for strong nanoscale variations in the bulk electronic properties includes: (1) a low temperature quasiparticle lifetime at least an order of magnitude shorter[24] in Bi-2212 than in $YBa_2Cu_3O_{7-\delta}$, (2) optical conductivity measurements demonstrating a finite low-temperature $\sigma_1$ (ref. 24) that can be explained by strong nanoscale variations in local superfluid density[25], (3) heat capacity measurements showing that with underdoping, $\boldsymbol{\gamma} = C/T = dS/dT$ falls rapidly and the temperature range of the transition widens dramatically[26], (4) NMR studies showing very broad line-widths and additional spin susceptibility at low-temperature indicative of strong electronic disorder[27], (5) inelastic neutron scattering line-widths that are significantly wider in Bi-2212 than in other cuprates[28], (6) ARPES studies showing similar evolution of spectra with deliberately introduced bulk disorder as with underdoping[29], and (7) interlayer tunneling experiments demonstrating the co-existence of superconducting and non-superconducting energy gaps[30,31]. Results from these different techniques would be consistent with each other and with STM data, if nanoscale variations exist in the properties of *bulk* Bi-2212, and if they are amplified by underdoping.

While the data in Figs. 1-3 are quite suggestive of granular superconductivity, single-particle excitation spectra cannot definitively distinguish between this scenario and a highly disordered single-phase superconductor. However, a novel atomic-scale technique, designed to locally distinguish superconducting from non-superconducting regions by using quasiparticle scattering-resonances at Ni impurity atoms, has recently been proposed[7-9]. A scattering-resonance originates from interactions between quasiparticles and an impurity atom and appears as an additional peak(s) in the excitation spectrum near the impurity. Ni scattering-resonances are particle-hole symmetric in the superconducting state[6], are predicted to lose their particle-hole symmetry if an electronic 'pseudogap' is present[8,9], and might disappear completely if conventional quasiparticles are nonexistent[7]. Thus, study of Ni scattering-resonances as a function of local-Δ should, in theory, allow the discrimination of superconducting from non-superconducting regions.

We apply this technique by carrying out spectral-surveys on as-grown Ni-doped Bi-2212 samples. Figure 4 summarizes results from two such experiments. Figure 4a shows the positions of the Ni scattering-resonances (which are identified by their strongest resonance peak near +18 meV [6]) superimposed on the gapmap derived from the same spectral-survey. The color scale is such that regions with Δ<50 meV (light gray) are distinguished from those with $\Delta>50$ meV (dark gray). Remarkably, no Ni scattering-resonances are observed in any region where $\Delta>50 \pm 2.5$ meV. To clarify the correlation between Ni- resonances and $\Delta$, we plot in Fig. 4b a combined histogram of observed Ni-resonances versus local-$\Delta$ from two independent spectral-surveys on different crystals. The combined histogram of $\Delta$ from the same two spectral-surveys is also shown in gray. Although for $\Delta<35$meV the two distributions are similar in shape, above $\Delta\sim 35$ meV they rapidly diverge and the Ni-resonance distribution reaches zero by 50meV.

There are several possible explanations for the Ni-resonance distribution. First, the amplitude of the 18meV resonance peak might fall with rising $\Delta$ and disappear near $\Delta\sim 50$meV. However, measurement of the peak heights shows them to be almost independent of local-$\Delta$. Second, there might be statistical fluctuations such that no Ni atoms reside in regions where $\Delta>50$meV. However, if the Ni atoms are distributed randomly, and if all regions can support quasiparticle scattering resonances, our non-observation (in two independent experiments on different crystals) of Ni-resonances in regions with $\Delta>50$meV has a combined probability of no more than $3\times 10^{-5}$. Therefore this explanation also appears ruled out. A third explanation could be that the Ni atoms somehow seed nanoscale regions (perhaps by attracting the dopant oxygen atoms or the holes), influencing them to develop into superconducting domains with $\Delta<50$meV. This scenario seems unlikely because, in one Zn-doped sample studied, Zn impurity-resonances disappear in a statistically similar fashion near $\Delta\sim 50$meV, but it cannot be ruled out. Nevertheless, impurity atoms are clearly *not* necessary to create $\alpha$-domains since they exist in samples with no impurity atoms (see Figs. 1a, 2,&3).

A final hypothesis is that Ni impurity atoms *are* physically present in regions with $\Delta>50$meV, but that they do not create scattering-resonances because these regions represent an electronically distinct phase. If Ni atoms are randomly distributed, and if the locations of particle-hole symmetric Ni-resonances indicate the local existence of superconductivity, then the distribution of superconducting regions has a similar shape to the red histogram of Fig. 4b. The picture would then be of purely superconducting regions when $\Delta\lesssim 35$ meV, a mixture of two different electronic orders when $35\text{meV}\lesssim\Delta\lesssim 50\text{meV}$, and an unidentified second phase (possibly the pseudogap) when $\Delta\gtrsim 50$meV. The data in Figs. 1-3 corroborates this picture particularly because the energy where the superconducting $\alpha$-domains disappear is identical to the energy where the Ni-resonances disappear. Therefore, although we cannot distinguish between the possible microscopic mechanisms[2,4,5,15-22] for the phenomena reported here, the data all suggest that underdoped Bi-2212 is a granular superconductor. This provides a new and unconventional context in which to view the underdoped cuprates.

We acknowledge A.V. Balatsky, E. Dagotto, M.E. Flatté, S.A. Kivelson, V.Z. Kresin, R.B. Laughlin, J.W. Loram, D.H. Lee, P.A. Lee, I. Martin, D.K. Morr, D. Pines, D.J. Scalapino, Z.-X. Shen, N. Trivedi, and S.A. Wolf for helpful discussions and communications. This work was supported by the LDRD program of Lawrence Berkeley National Laboratory, by the ONR, by the CULAR program of Los Alamos National Laboratory, by the Miller Research Foundation (JCD), by IBM (KML); and by Grant-in-Aid for Scientific Research on Priority Area (Japan), a COE Grant from the Ministry of Education.

**Correspondence and requests for materials should be addressed to J.C.D. (email: jcdavis@socrates.berkeley.edu).**

**Figure 1** Typical 560 Å square gapmaps from 'underdoped' and 'as-grown' Bi-2212 samples.

We use an identical color-scale to represent $\Delta$ in these two panels for ease of comparison. The color-scale spans the range 20meV-64meV, and ticks on the scale are placed at 34meV, 42meV, and 50 meV. These gapmaps were calculated from two 128x128 pixel spectral-surveys. **a**. Gapmap of an oxygen-underdoped Bi-2212 sample with $T_c \approx$79 K. This field of view contains approximately 130 compact low-$\Delta$ domains which are visible as orange, yellow and green islands against the interconnected blue and black background which are the high-$\Delta$ regions. **b**. Gapmap of an as-grown Bi-2212 sample. In this figure the interconnected red, orange, and yellow low-$\Delta$ regions completely dominate the image. Embedded in this environment are filamentary blue and black high-$\Delta$ regions. Note that although this sample contains a 0.5% substitution of Ni for the Cu atoms in the $CuO_2$ plane, the 102 impurity resonances in this field of view affect less than 15% of the area, and furthermore, the local value of $\Delta$ does not vary near the Ni impurities[6]. Both spectral-surveys were acquired with constant current normalization in cryogenic UHV at 4.2K with tunnel junction resistance of 1G$\Omega$ set at V = –100mV.

**Figure 2** High spatial resolution data showing the typical spatial interrelationship of $\Delta$ and $G(\Delta)$ for underdoped Bi-2212.

The data in this figure were generated from a single 128x128 pixel spectral-survey which was taken on the 147 Å square region inside the white box in Fig 1a. This spectral-survey was taken several weeks after the survey used to calculate figure Fig. 1a, during which the sample was kept continuously at 4.2 K. **a**. High resolution gapmap revealing approximately 12 $\alpha$-domains embedded in the percolative $\beta$-like background. **b**. Map of $G(\Delta)$ at the same location as **a**. Here, to reveal the spatial correlations between $\Delta$ and $G(\Delta)$, we use the same color-bar as in **a** but here the scale is inverted and represents $G(\Delta)$. Note that $G(\Delta)$ is defined as the gap-edge peak at negative bias throughout this paper. In **c** we show the spatial evolution of $\Delta$ and in **d** of $G(\Delta)$ along the trajectory of the white line seen in **a** and **b**. Figures 2 and 3 together illustrate the properties of the two distinct types of regions. In terms of $\Delta$, the $\alpha$-domains are defined by a distinct value of $\Delta$ which is $\lesssim$50meV and which is constant to 5% within a 13Å radius of the center of each domain, while the $\beta$-regions are defined by $\Delta \gtrsim$35meV. There exists an overlap range of $\Delta$ (35meV$\lesssim\Delta\lesssim$50meV) in which categorization must rely on parameters in addition to $\Delta$. The tick mark on the color-scale is at 42meV at the center of this overlap range. In terms of $G(\Delta)$, the $\alpha$-domains are defined by $G(\Delta)$ at the center of the domain $\approx$66% ($\sigma$=23%) above the average $\beta$-region $G(\Delta)$, which itself has $\sigma$=15%. Similar conclusions are derived from all equivalent maps from other underdoped spectral-surveys. The spectral-survey was acquired with constant current normalization in cryogenic UHV at 4.2K with tunnel junction resistance of 1GΩ set at V = –200mV.

**Figure 3** Typical series of dI/dV spectra illustrating how the two distinct types of regions are apparent in the raw data.

**a**. These unprocessed spectra were extracted from the same spectral-survey used to create Fig.2, along the white line shown in Fig. 2a&b. The spectra are separated by 1.1Å and span a total distance of 140 Å as represented by their vertical offset. The α-domain spectra (red) have low gap magnitudes and sharp gap-edge peaks whose amplitude is low at the edges of the domain and rises to a maximum in the center. The β-region spectra have high gap magnitudes and very broad gap-edge peaks whose amplitude is relatively low and constant. An additional feature seen in this figure are the spectra which lie on the borders between neighboring distinct regions. These spectra can show 4 peaks, where each pair of peaks corresponds to the gap value of the neighboring regions. For the maps shown in Figs. 1&2, the peak with the overall maximum value of G(eV) is chosen as G(Δ). In Figs. 2c&d, if four clear peaks can be identified in the dI/dV spectrum, then both values of Δ and G(Δ) at that location are plotted. **b**. Surface topography with the trajectory along which both the spectra in **a,** and the traces of 2c&d, were measured, shown as a white line. The atomic resolution associated with the spectral-surveys is clearly evident.

**Figure 4** Analysis of the relationship between Ni scattering-resonances and local-Δ.

Spectral-surveys were acquired at 4.2K on two as-grown Ni-doped Bi-2212 samples, both with tunnel junction resistance of 1GΩ set at V=–100mV. Using the +18mV LDOS image of each survey, the locations of Ni resonances were identified by the strong resonance peak at +18 mV in their dI/dV spectra[6]. For ≈30% of these resonances the full particle-hole symmetric signature (spatial and spectroscopic) of a Ni-resonance[6] was directly confirmed. The location of the Ni resonances can be correlated with the local value of Δ extracted from these same two surveys. **a**. shows the location of the Ni resonances as red dots superimposed on the simultaneously acquired gapmap of area 600Å square. No Ni resonances are observed in regions where Δ>50 meV. **b**. shows histograms derived from a combined analysis of the above spectral-survey on 0.2% Ni-doped Bi-2212, and an independent 568Å survey on 0.5% Ni-doped Bi-2212. For each Ni resonance, the local-Δ is determined by the spatial average of Δ over a 13Å square region centered on the impurity. Using this information, we plot in red a combined histogram of observed Ni resonances versus Δ from both samples. The combined histogram of Δ from the same two surveys is shown in gray.

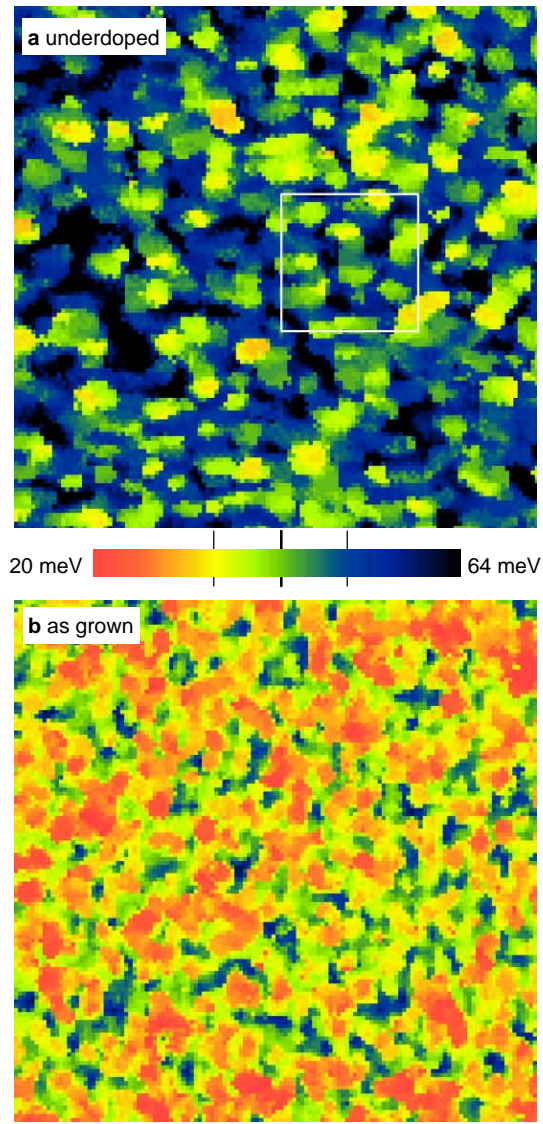

K.M. Lang, *et al.*
Figure 1

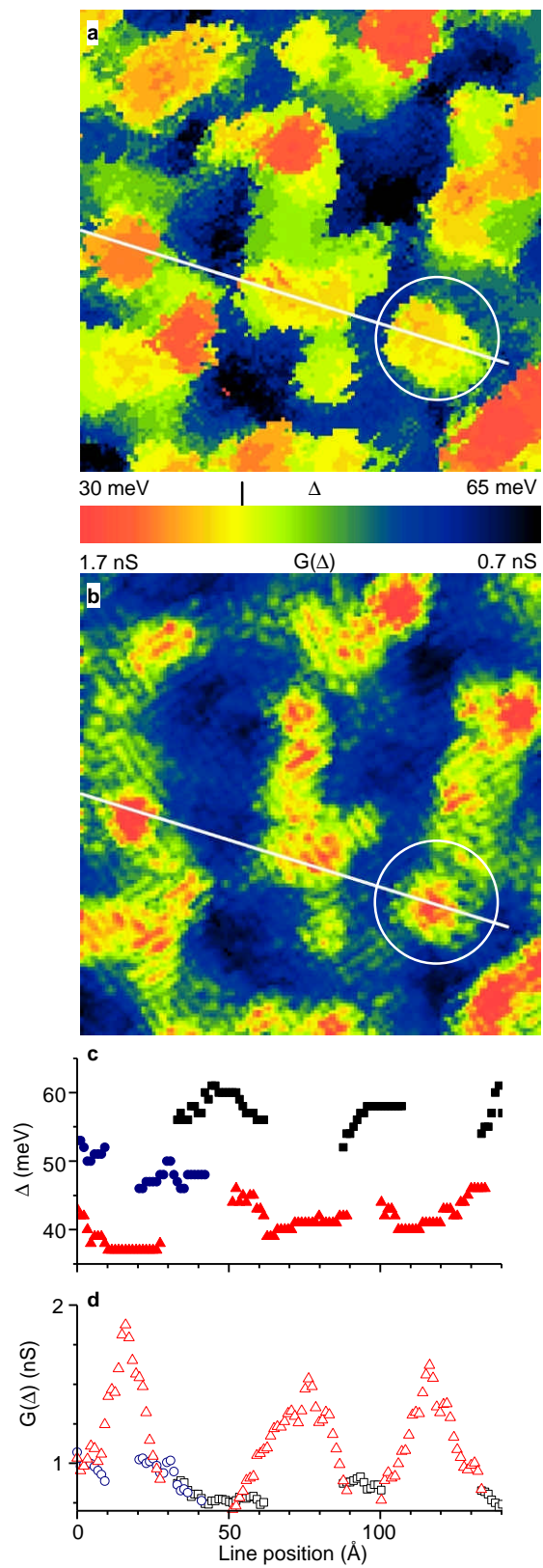

K.M. Lang, *et al.*
Figure 2

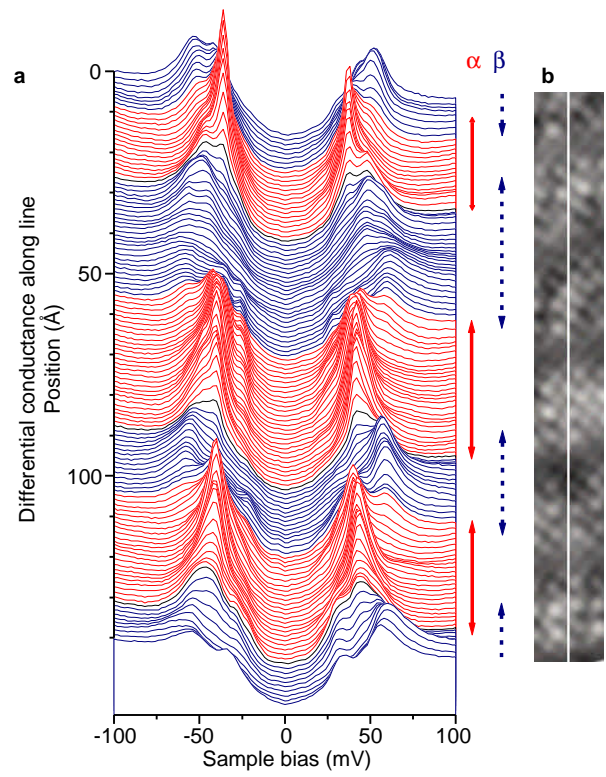

K.M. Lang, *et al.*
Figure 3

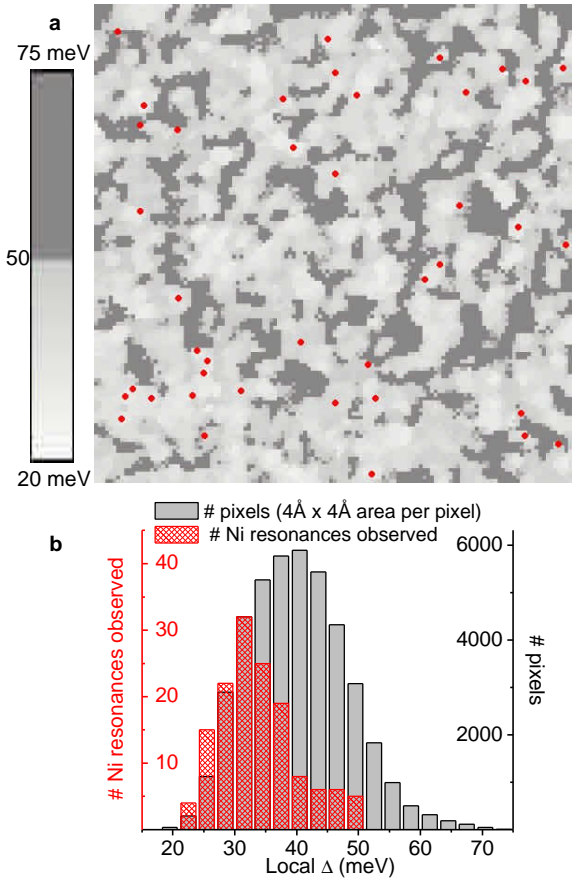